\documentclass[12pt]{article}

\usepackage{amsmath,amsthm,amssymb,a4wide}

\usepackage{mybbold}    

\newcommand{\ud}{\text{d}}
\newcommand{\ui}{\text{i}}
\newcommand{\ue}{\text{e}}
\newcommand{\skl}{\mathrm{S}}
\newcommand{\scl}{\mathrm{S}_{\mathrm{cl}}}
\newcommand{\SU}{\mathrm{SU}}
\newcommand{\U}{\mathrm{U}}
\newcommand{\su}{\mathrm{su}}
\newcommand{\SO}{\mathrm{SO}}
\newcommand{\Ad}{\mathrm{Ad}}

\newcommand{\al}{\alpha}
\newcommand{\be}{\beta}
\newcommand{\de}{\delta}
\newcommand{\Ga}{\Gamma}
\newcommand{\la}{\lambda}
\newcommand{\om}{\omega}
\newcommand{\Om}{\Omega}
\newcommand{\si}{\sigma}
\newcommand{\Si}{\Sigma}
\newcommand{\ve}{\varepsilon}
\newcommand{\vp}{\varphi}

\newcommand{\cD}{{\mathcal D}}
\newcommand{\cL}{{\mathcal L}}
\newcommand{\cM}{{\mathcal M}}
\newcommand{\cS}{{\mathcal S}}

\newcommand{\vecA}{\boldsymbol{A}}
\newcommand{\vecB}{\boldsymbol{B}}
\newcommand{\vecb}{\boldsymbol{b}}
\newcommand{\vecC}{\boldsymbol{C}}
\newcommand{\vecv}{\boldsymbol{v}}
\newcommand{\vecw}{\boldsymbol{w}}
\newcommand{\vecx}{\boldsymbol{x}}
\newcommand{\vecy}{\boldsymbol{y}}
\newcommand{\vecp}{\boldsymbol{p}}
\newcommand{\vecq}{\boldsymbol{q}}
\newcommand{\vecsig}{\boldsymbol{\sigma}} 
\newcommand{\vecSig}{\boldsymbol{\Sigma}}
\newcommand{\vecnab}{\boldsymbol{\nabla}}
\newcommand{\vecxi}{\boldsymbol{\xi}}

\newcommand{\rz}{{\mathbb R}}
\newcommand{\nz}{{\mathbb N}}
\newcommand{\kz}{{\mathbb C}}

\DeclareMathOperator{\tr}{Tr}
\DeclareMathOperator{\mtr}{tr}
\DeclareMathOperator{\vol}{vol}         

\newcommand{\eins}{\mathmybb{1}}

\numberwithin{equation}{section}

\newtheorem{theorem}{Theorem}[section]
\newtheorem{lemma}[theorem]{Lemma}
\newtheorem{prop}[theorem]{Proposition}

\theoremstyle{definition}
\newtheorem{defn}[theorem]{Definition}

\begin{document}

\bibliographystyle{amsalpha}

\thispagestyle{empty}

\noindent
ULM-TP/00-1\\
April 2000\\

\vspace*{1cm}

\begin{center}

{\LARGE\bf Quantum ergodicity for Pauli Hamiltonians\\
\vspace*{3mm}
with spin 1/2} \\
\vspace*{2cm}
{\large Jens Bolte}%
\footnote{E-mail address: {\tt bol@physik.uni-ulm.de}} 
{\large and Rainer Glaser}%
\footnote{E-mail address: {\tt gla@physik.uni-ulm.de}}\\ 

\vspace*{1cm}

Abteilung Theoretische Physik\\
Universit\"at Ulm, Albert-Einstein-Allee 11\\
D-89069 Ulm, Germany 
\end{center}

\vfill

\begin{abstract}
Quantum ergodicity, which expresses the semiclassical convergence of almost 
all expectation values of observables in eigenstates of the quantum 
Hamiltonian to the corresponding classical microcanonical average, is proven 
for non-relativistic quantum particles with spin 1/2. It is shown that quantum
ergodicity holds, if a suitable combination of the classical translational
dynamics and the spin dynamics along the trajectories of the translational
motion is ergodic.
\end{abstract}

\newpage

\setcounter{footnote}{0}

\section{Introduction}
\label{sec1}

In quantum chaos one is primarily interested in statistical properties
of eigenvalues and eigenvectors of quantum Hamiltonians whose classical
limits generate chaotic dynamics. In this context the eigenvalue statistics
on the scale of the mean level spacing are expected to be described by
random matrix theory (see, e.g., \cite{BohGiaSch84,Meh91}). This
conjecture has since found overwhelming confirmation, mostly based on 
numerical calculations of eigenvalues in many systems. 
For the corresponding eigenvectors one also expects a random behaviour,
and various tools to measure this have been invented. Among them one
finds one of the few results in quantum chaos for which a mathematical 
proof is available: if a classical system is ergodic, its quantum mechanical 
counterpart is quantum ergodic. By this one understands a semiclassical
convergence of almost all phase-space lifts of eigenfunctions 
of the Hamiltonian (e.g., their Wigner- and Husimitransforms) towards 
equidistribution with respect to microcanonical (i.e.\ Liouville) measure.
One thus has obtained a realisation of the semiclassical eigenfunction
hypothesis (see, e.g., \cite{Ber83}) for classically ergodic systems.

Quantum ergodicity was first established for the free motion of a particle 
on a compact Riemannian manifold, where the quantum dynamics is generated 
by minus the Laplace-Beltrami operator on that manifold. This goes back 
to Shnirelman \cite{Shn74}, and the first complete proofs are due to 
Zelditch \cite{Zel87} and Colin de Verdi\`ere \cite{Col85}. In the systems
considered by these authors the semiclassical limit is actually realised 
as a high-energy limit. However, for Schr\"odinger operators involving a 
potential and possibly also a magnetic field the semiclassical limit can 
in general only be performed in terms of $\hbar\to 0$. In this setting 
quantum ergodicity was first proven by Helffer, Martinez, and Robert 
\cite{HelMarRob87}. All of the work mentioned so far exclusively dealt with 
the dynamics of point particles without internal degrees of freedom; in 
particular no spin was involved. On the level of the quantum Hamiltonians 
$\hat H$ that were taken into account, this is reflected in the fact that 
$\hat H$ appears as a (Weyl-) quantisation of a scalar symbol. The only
degrees of freedom appearing are therefore those that possess a direct 
classical analogue. These are in fact the translational degrees of freedom, 
which on the classical side yield as a phase space a smooth symplectic
manifold that in typical cases is the cotangent bundle over the
configuration manifold. In this setting internal degrees of freedom would
appear through the fact that the quantum mechanical observables are
quantisations of matrix valued symbols, such that the vector spaces these
matrices operate on represent the internal degrees of freedom.

In quantum mechanics internal degrees of freedom often arise due to the
presence of symmetries. Since in general symmetries have to be implemented 
through (anti-) unitary representations of the respective
symmetry groups on the Hilbert space of state vectors, representations
of dimensions exceeding one introduce discrete degrees of freedom. It
then can happen that these additional degrees of freedom do not possess
classical analogues. A prominent example of this phenomenon is provided
by the spin of a particle, which arises through the space-time symmetries
that either form the Lorentz group (in a relativistic theory), or the
Galilei group (in a non-relativistic theory). Both these symmetry groups
contain the proper rotations, i.e.\ $\SO(3)$, as a subgroup. Through the 
unitary projective representations of the Lorentz or Galilei group,
respectively, which implement the space-time symmetries in the quantum 
theory, one therefore introduces the unitary irreducible representations of 
$\text{Spin}\,(3)=\SU(2)$, i.e.\ spin. On the classical side the rotation
group, however, remains to be $\SO(3)$, which then enters in terms of spatial 
angular momentum. There exists, however, no direct classical analogue of spin.
Coming back now to the problem of quantum ergodicity for particles with spin,
there immediately arises the question for a criterion to be imposed on the
classical system in order to ensure that the eigenfunctions of the
quantum Hamiltonian behave quantum ergodically in the semiclassical limit.
We recall that in the case without spin such a criterion was given by
the ergodicity of the dynamics on the classical phase space with respect
to Liouville measure. One would now expect that the presence of additional, 
internal degrees of freedom requires an extended criterion in order that 
quantum ergodicity holds. 

It is the primary goal of the present work to elaborate on this question in 
some detail for the case of a non-relativistic quantum particle with spin 
1/2. Our proof of quantum ergodicity in this context generalises the 
methods of \cite{Shn74,Zel87,Col85,HelMarRob87} to the situation of Weyl 
operators with $2\times 2$-matrix valued symbols. This requires two 
essential ingredients. The first one is an Egorov Theorem, which relates 
the semiclassical limit of the quantum mechanical time evolution of an 
observable to the classical time evolution of the corresponding classical 
observable. The second input required is a Szeg\"o limit formula, which 
expresses the semiclassical limit of averaged expectation values of 
observables in eigenstates of the Hamiltonian in terms of a classical 
microcanonical average. In the course of the subsequent proof of quantum 
ergodicity we primarily rely on the method developed in 
\cite{Zel96,ZelZwo96}, which provides a considerable simplification
over the original proofs given in \cite{Zel87,Col85,HelMarRob87}. Already
the Egorov property that one encounters for systems with spin hints at
the construction that yields the `classical' criterion for quantum
ergodicity that we seek for. It leads us to consider an $\SU(2)$-extension
of the Hamiltonian flow that arises from the classical limit of the 
translational degrees of freedom. The thus extended flow is defined on
a product phase space that consists of two parts: the translational part
is the hypersurface of fixed energy in the classical phase space,
whereas the spin part is given by the group manifold of $\SU(2)$. The
translational part of the combined dynamics is then provided by the
Hamiltonian flow. The latter also drives the spin dynamics that takes place 
on $\SU(2)$ and consists of a left multiplication by a spin transport  
matrix, which propagates the spin degrees of freedom along the trajectories
of the Hamiltonian flow. Our main result, stated in Theorem \ref{QE}, then 
is that ergodicity of the combined flow is a sufficient criterion for 
quantum ergodicity. At this point we remark that the problem of quantum 
ergodicity for quantisations of matrix valued symbols was already considered 
in \cite{Zel96} as an example for the general theory of quantum ergodicity
of $C^*$ dynamical systems. There it was, however, overlooked that in
the relevant Egorov Theorem, apart from a transport by the Hamiltonian
flow, the principal symbol of an observable is also conjugated with the
spin transport matrices. As a consequence, the result stated in \cite{Zel96}  
is hence incorrect, to the extent that ergodicity of the Hamiltonian
flow alone is insufficient to guarantee quantum ergodicity. In section
\ref{sec5} we give an example that illustrates this fact.

This paper is organised as follows. In section \ref{sec2} we review some
background on $\hbar$-pseudodifferential calculus and define the type
of Hamiltonians and observables that will be considered in the sequel. 
The two major ingredients required for quantum ergodicity, namely the
Egorov property and the Szeg\"o limit formula, are developed in section
\ref{sec3}. Our main result, quantum ergodicity for Pauli Hamiltonians
with spin 1/2, is then proven in section \ref{sec4}. Finally, in section
\ref{sec5} we discuss the consequences of quantum ergodicity for Wigner- and 
Husimitransforms of eigenfunctions and, furthermore, give an example that 
illustrates why ergodicity of the Hamiltonian flow alone is not a sufficient 
criterion for quantum ergodicity.

\section{Semiclassical background}
\label{sec2}

In non-relativistic quantum mechanics the dynamics of a particle with
spin $s\in\frac{1}{2}\nz$ is governed by the Pauli equation
\begin{equation} 
\label{Paulieq}
\ui\hbar\,\frac{\partial\psi}{\partial t} (\vecx, t) = 
\hat{H}_P \psi(\vecx,t)
\end{equation} 
with the quantum Hamiltonian
\begin{equation} 
\label{Pauliop}
\hat{H}_P = \hat{H}_{trans.}\eins_{2s+1} + \hbar\,\vecSig\cdot\hat{\vecC}
\end{equation}
acting as a self-adjoint operator on a suitable domain in the Hilbert space 
$L^2(\rz^3)\otimes\kz^{2s+1}$. Here 
\begin{equation} 
\label{Htrans}
\hat{H}_{trans.} = \frac{1}{2m} \left( \frac{\hbar}{\ui} \vecnab_x - 
\frac{e}{c} \vecA(\vecx) \right)^2 + e \vp(\vecx)
\end{equation}
describes the dynamics of the translational degrees of freedom of a 
spinning particle with mass $m$ and charge $e$ which is subject to external 
electromagnetic forces generated by the (static) potentials $\vecA$ and 
$\vp$. Furthermore, the components $\Si_k$, $k=1,2,3$, of $\vecSig$ denote 
the hermitian generators of the Lie algebra $\su(2)$ in the (irreducible) 
spin-$s$ representation, which is of dimension $2s+1$. The coupling of the 
spin degrees of freedom to the translational ones is provided by the 
operators $\hat C_k$, which are suitable quantisations of functions 
$C_k(\vecp,\vecx)$ on phase space. 
The latter can, e.g., describe a coupling to an external magnetic field, 
$\vecC_B(\vecp,\vecx)=-\frac{e}{2mc}\vecB(\vecx)$, or a spin-orbit coupling 
$\vecC_{\text{so}}(\vecp,\vecx) = \frac{1}{4m^2 c^2 |\vecx|}
\frac{\ud\vp(|\vecx|)}{\ud |\vecx|}(\vecx\times\vecp)$. 

Here, and in the following, we choose all quantum mechanical observables 
to be Weyl quantisations of matrix valued symbols. In general this is
defined for $B\in\cS'(\rz^d\times\rz^d)\otimes\kz^{n\times n}$ and 
$\psi\in\cS(\rz^d)\otimes\kz^n$ as
\begin{equation}
\label{Weylq} 
(\hat{B}\psi)(x) := \frac{1}{(2\pi\hbar)^d} \int_{\rz^d} \int_{\rz^d}
\ue^{\frac{\ui}{\hbar}\vecp\cdot(\vecx-\vecy)}\,B\Bigl( \vecp,\frac{1}{2}
\bigl(\vecx+\vecy\bigr)\Bigr)\,\psi(\vecy)\ \ud y\,\ud p \ ,
\end{equation} 
where $\hbar\in (0,\hbar_0]$ serves as a parameter. The quantisation
(\ref{Weylq}) yields a continuous map from $\cS(\rz^d)\otimes\kz^n$ 
to $\cS'(\rz^d)\otimes\kz^n$. However, in order to obtain a semiclassical 
calculus one has to restrict attention to smaller classes of symbols and 
operators. In particular, one wishes to consider operators that can be 
composed with one another, e.g., operators which map $\cS(\rz^d)\otimes\kz^n$ 
into itself. In defining suitable symbol classes we use the following notion, 
which is in accordance with \cite{DimSjo99}: 
 
A function $m:\rz^d\times\rz^d\to (0,\infty)$ is called an 
\emph{order function}, if there are constants $C_0 >0, \; N_0 >0$ such that 
\begin{equation}
m(\vecp,\vecx) \le C_0 \sqrt{ 1+ (\vecp-\vecq)^2 + (\vecx-\vecy)^2}^{N_0} \ 
m(\vecq,\vecy)\ .
\end{equation}
An example for such an order function is 
\begin{equation}
m(\vecp,\vecx) = 1+\vecp^2 +\vecx^2 \ .
\end{equation}
\begin{defn} Let $m$ be an order function on $\rz^d\times\rz^d$. We define 
the symbol class $\skl(m)$ to be the set of $B\in C^\infty(\rz^d\times\rz^d)
\otimes\kz^{n\times n}$ such that for every $\al,\be\in\nz^{d}_0$ there 
exists $C_{\al,\be}>0$ with
\begin{equation}
\| \partial^\al_p \partial_x^\be B(\vecp,\vecx) \| \le C_{\al,\be} 
\,m(\vecp,\vecx)\ , 
\end{equation}
where $\| \cdot \|$ is some (arbitrary) matrix norm on $\kz^{n\times n}$. 
\end{defn}
If $B=B(\vecp,\vecx ;\hbar)$ depends on $\hbar\in (0,1]$, we say that 
$B\in\skl(m)$, if $B(\cdot,\cdot;\hbar)$ is uniformly bounded in $\skl(m)$ 
when $\hbar$ varies in $(0,1]$. For $k\in\rz$ we let $\skl^k(m)$ be the set 
of functions $B(\vecp,\vecx;\hbar)$ on $\rz^d\times\rz^d\times (0,1]$ 
that belong to $\hbar^{-k}\skl(m)$ and satisfy
\begin{equation}
\| \partial^\al_p \partial_x^\be B(\vecp,\vecx;\hbar) \| \le C_{\al,\be} 
\,m(\vecp,\vecx)\,\hbar^{-k}\ .
\end{equation} 
A sequence of symbols $B_j\in\skl^{k_j}(m)$, with $k_j\to -\infty$ 
monotonically, defines an asymptotic expansion of $B\in\skl^{k_0}(m)$, 
denoted by
\begin{equation} 
\label{eq:classic_expansion}
B\sim\sum_{j=0}^\infty B_j \ ,
\end{equation}
if
\begin{equation}
B - \sum_{j=0}^N B_j \in \skl^{k_{N+1}}(m) 
\end{equation}
for every $N\in\nz_0$.
We will often use the smaller class of \emph{classical symbols} $\scl^k(m)$, 
whose elements $B\in\scl^k(m)$ possess asymptotic expansions in integer
powers of $\hbar$, i.e.
\begin{equation}
\label{asymcl}
B\sim\sum_{j=0}^\infty \hbar^{-k+j} B_j\ , \qquad B_j\in\skl(m)\ .
\end{equation}
In all of the above symbol classes the composition of the corresponding
Weyl operators is well defined in the following sense (see, e.g., 
\cite{DimSjo99}):
\begin{lemma} 
\label{lem:produktformel}
Let $m_1,m_2$ be order functions. For $B_j\in\skl(m_j)$ the product
of the associated Weyl operators reads in terms of their symbols
\begin{equation}
\hat{B}_1\hat{B}_2 = \widehat{B_1 \# B_2}\ ,
\end{equation}
where $(B_1,B_2)\mapsto B_1 \# B_2$ is a bilinear continuous map from 
$\skl(m_1) \times \skl(m_2)$ to $\skl(m_1 m_2)$. It is explicitly given 
by 
\begin{equation}
\label{symbprod} 
(B_1 \# B_2)(\vecp,\vecx) = \left. \ue^{ \frac{\ui\hbar}{2} 
\si(\partial_p,\partial_x;\partial_q,\partial_y)}\, B_1(\vecp,\vecx)\,
B_2(\vecq,\vecy) \right|_{\substack{\vecq=\vecp \\ \vecy=\vecx}} \ ,
\end{equation}
where $\sigma(\vecv_p,\vecv_x;\vecw_p,\vecw_x):=\vecv_x\cdot\vecw_p - 
\vecv_p\cdot\vecw_x$ denotes the symplectic two-form on $\rz^d\times\rz^d$. 
Furthermore, the asymptotic expansion of (\ref{symbprod}) in $\skl(m_1 m_2)$ 
reads
\begin{equation}
\label{prodform} 
(B_1 \# B_2)(\vecp,\vecx) \sim \left. 
\sum_{k=0}^\infty \frac{1}{k!} \left( \frac{\ui \hbar}{2} 
\si(\partial_p,\partial_x;\partial_q,\partial_y) \right)^k \,
B_1(\vecp,\vecx)\,B_2(\vecq,\vecy) \right|_{\substack{\vecq=\vecp \\ 
\vecy=\vecx}}\ .
\end{equation}
\end{lemma}
The Pauli Hamiltonians (\ref{Pauliop}) that we are going to study below
can be viewed as Weyl quantisations of hermitian symbols $H\in\scl^0(m)$, 
where $m$ is an order function with $m\ge 1$. The symbol $H$ then has an 
asymptotic expansion of the type (\ref{asymcl}),
\begin{equation}
H \sim \sum_{j=0}^\infty \hbar^j H_j\ , \qquad H_j \in\skl(m)\ ,
\end{equation}
and the principal symbol $H_0$ is supposed to be a scalar multiple of 
$\eins_n$, i.e.\ $H_0 =H_{0,s}\eins_n$. Since in quantum mechanics Hamilton 
operators must be self-adjoint and bounded from below, we require the real 
valued function $H_{0,s}$ to fulfill the following properties:
\begin{enumerate}
\item $H_{0,s}$ is bounded from below.
\item There exists an energy value $E\in\rz$ and some $\ve>0$ such 
that $H_{0,s}^{-1}([E-\ve,E+\ve])\subset\rz^d\times\rz^d$ is compact.
\item $H_{0,s}$ has no critical value in $[E-\ve,E+\ve]$.
\item $(H_{0,s}+\ui)$ is elliptic in the sense that
\begin{equation} 
\label{eq:ellipticity}
|H_{0,s}(\vecp,\vecx)+\ui| \ge C\,m(\vecp,\vecx)\ ,
\end{equation}
with some $C>0$.
\end{enumerate}
If $\hbar$ is small enough, the above properties of $H_{0,s}$ ensure that 
the spectrum of $\hat{H}$ is purely discrete in any interval which is 
properly contained in $[E-\varepsilon,E+\varepsilon]$, and that $\hat H$ 
is essentially self-adjoint on $C^\infty_0 (\rz^d)\otimes\kz^n$. For 
simplicity, we denote the
self-adjoint extensions, which are the relevant quantum mechanical 
Hamiltonians, also by $\hat H$. Furthermore, if $f\in C_0^\infty(\rz)$, 
the operator $f(\hat{H})$ defined by the functional calculus given by the 
spectral theorem is a Weyl operator with symbol $\phi\in\skl^0(m^{-r})$ 
for every $r\in\rz$. Its asymptotic expansion is given by
\begin{equation}
\phi \sim \sum_{j=0}^\infty \hbar^{j} \phi_j\ ,
\end{equation}
with $\phi_0(\vecp,\vecx)=f(H_{0,s}(\vecp,\vecx))\eins_n$ and 
$\phi_1(\vecp,\vecx)=H_1(\vecp,\vecx)f'(H_{0,s}(\vecp,\vecx))$. The above
properties are proven in \cite{Rob87,DimSjo99} for the case
of scalar valued symbols. These proofs can immediately be carried over to
the present situation in which the principal symbol $H_0$ of the Hamiltonian
is a scalar multiple of $\eins_n$.

We remark that the requirement $H_0\in\skl(m)$ together with 
(\ref{eq:ellipticity}) implies that 
\begin{equation}
C^2\,m^2(\vecp,\vecx) \le 1+H_{0,s}^2(\vecp,\vecx) \le 
D^2\,m^2(\vecp,\vecx)+1 \ , 
\end{equation}
so that $(1+H_{0,s}^2)^{1/2}$ is an order function. Moreover, the condition 
$H_j\in\skl (m)$ is equivalent to $H_j\in\skl((1+H_{0,s}^2)^{1/2})$, which 
is the requirement on the symbol imposed in 
\cite{Rob87,HelMarRob87}.

Below we will study the semiclassical behaviour of quantum mechanical
observables and of their expectation values in eigenstates of the
Hamiltonian. For these purposes it will be advantageous to restrict
attention to bounded observables. In order to characterise a sufficiently
large class of such operators we employ
\begin{prop}[Calder\'on-Vaillancourt]
Let $B\in\skl^0(1)$, then the Weyl quantised operator $\hat B$ is bounded
on $L^2(\rz^d) \otimes \kz^n$.
Moreover, for $\hbar\in (0,1]$ there exists an upper bound for the 
operator norms of $\hat B$.
\end{prop}
The original version of this Proposition goes back to Calder\'on and 
Vaillancourt \cite{CalVai71}. In the form presented here, it can be found
(for scalar valued symbols) in \cite{DimSjo99}; that proof can be directly 
carried over to the present situation. Furthermore,
if we require the symbols $B\in\skl^0(1)$ to be hermitian matrices,
their Weyl quantisations $\hat B$ are symmetric operators that can be
extended to self-adjoint operators on all of $L^2(\rz^d)\otimes\kz^n$; 
the latter we also denote by $\hat B$. The above properties can immediately
be generalised to the case $B\in\skl^k(1)$. Then, however, one has to
extract a factor of $\hbar^{-k}$ from the norm of $\hat B$ in order to 
obtain the bound on the norms when $\hbar$ varies in $(0,1]$.

\section{Semiclassical time evolution and Szeg\"o limit formula}
\label{sec3}

The quantum mechanical time evolution as, e.g., governed by the Pauli
equation (\ref{Paulieq}) requires to investigate the Cauchy problem for 
the operator
\begin{equation}
\ui\hbar\,\frac{\partial}{\partial t} - \hat{H}
\end{equation}
on $L^2(\rz^d)\otimes\kz^n$. This problem can be solved by introducing the 
one-parameter group of unitary operators
\begin{equation}
\hat{U}(t):= \ue^{-\frac{\ui}{\hbar}\hat{H}t}\ ,
\end{equation}
which are well defined for all $t\in\rz$ since $\hat{H}$ is (essentially) 
self-adjoint. The time evolution of a quantum mechanical observable 
$\hat{B}$, which arises as the Weyl quantisation of a symbol 
$B\in\scl^k (1)$, is then given by
\begin{equation}
\label{Boft}
\hat{B}(t):= \hat{U}^\dagger(t)\,\hat{B}\,\hat{U}(t)\ .
\end{equation}
The time-evolved observable (\ref{Boft}) therefore satisfies the Heisenberg 
equation of motion
\begin{equation} 
\label{eq:heisenberg-gleichung}
\frac{\partial}{\partial t}\,\hat{B}(t) = \frac{\ui}{\hbar}\,
[\hat{H},\hat{B}(t)] \ , \qquad \hat{B}(0)= \hat{B}\ .
\end{equation}
For Weyl operators $\hat{H}$ and $\hat{B}$ with scalar symbols it is well 
known that for finite times $t$ the propagators $\hat{U}(t)$ are 
semiclassical Fourier integral operators which arise as quantisations of 
the Hamiltonian flow generated by the principal symbol $H_{0,s}$ of the 
quantum Hamiltonian $\hat{H}$. In addition, the time evolution (\ref{Boft})
respects operator classes in the sense that if $\hat B$ is a quantisation
of a symbol in some suitable class such as $\scl^k(1)$, $\hat B(t)$ is
again an operator with symbol in the same class (see, e.g., \cite{Rob87}). 

In the case of matrix valued symbols the situation is different; in physical
terms this has to do with the need to propagate the internal (i.e.\ spin) 
degrees of freedom in addition to the translational ones. To begin with, 
let us hence introduce 
the Hamiltonian vector field $X_{H_{0,s}}:=(-\partial_x H_{0,s},\partial_p 
H_{0,s})$ associated with the scalar factor $H_{0,s}$ of the principal
symbol $H_0$, and denote by $\Phi^t$ the flow generated by $X_{H_{0,s}}$.
That is, $(\vecp(t),\vecx(t))=\Phi^t(\vecp,\vecx)$ is a solution of 
Hamilton's equations of motion
\begin{equation} 
\label{Hameqmo}
\dot{\vecp}(t) = -\partial_x H_{0,s}(\vecp(t),\vecx(t))\ , \qquad
\dot{\vecx}(t) = \partial_p H_{0,s}(\vecp(t),\vecx(t))\ ,
\end{equation}
with initial condition $(\vecp(0),\vecx(0))=(\vecp,\vecx)$. This describes
the classical dynamics of the translational degrees of freedom, whereas
the propagation of the spin degrees of freedom is only contained in
the dynamics of the (matrix valued) symbol $B(t)$. For the following
we suppose that $\hat B$ is a Weyl operator with symbol $B\in\scl^k(1)$.
Then (\ref{Boft}) yields the observable $\hat B (t)$, which is a Weyl
quantisation of some symbol $B (t)$ that in general will not be in
the class $\scl^k(1)$, although, however, $\hat B (t)$ clearly remains a 
bounded operator. We hence base the following construction on the formal
asymptotic expansion
\begin{equation}
\label{Boftasym}
B(t) \sim \sum_{l=0}^\infty \hbar^{-k+l}\,B_l(t) \ ,
\end{equation}
whose coefficients can be determined from the Heisenberg equations of
motion (\ref{eq:heisenberg-gleichung}), once these have been transfered
to the level of symbols with the help of the product formula 
(\ref{prodform}); the latter also applies to operators with symbols that 
have a formal asymptotic expansion of the type (\ref{Boftasym}). We thus
obtain the recursive Cauchy problem, 
\begin{multline} 
\label{eq:cauchy_problems}
\frac{\partial}{\partial t}B_l(t) - \{ H_0,B_l(t) \} - \ui\,[H_1,B_l(t)] = \\ 
\sum_{\substack{0\le k\le l-1 \\ j+|\al| + |\be|= l -k}} 
\frac{\ui^{|\al|-|\be|}}{2^{|\al|+|\be|} |\al|! |\be|!} \biggl( 
\bigl( \partial_p^\be \partial_x^\al B_k(t) \bigr) 
\bigl( \partial_p^\al \partial_x^\be H_j \bigr) - (-1)^{|\al|-|\be|} 
\bigl( \partial_p^\al \partial_x^\be H_j \bigr) 
\bigl( \partial_p^\be \partial_x^\al B_k(t) \bigr) \biggr)\ ,
\end{multline}
with $B_l(0) = B_l$, compare \cite[ch.2.3]{Ivr98}. Here
\begin{equation} 
\{ A,B \} := \partial_p A\cdot\partial_x B - \partial_x A\cdot\partial_p B
\end{equation}
denotes the Poisson bracket of $A,B\in C^{\infty}(\rz^d\times\rz^d)\otimes
\kz^{n\times n}$; notice that in general $\{ A,B \}\neq -\{ B,A \}$. In 
leading semiclassical order (\ref{eq:cauchy_problems}) now yields the 
following equation for the principal symbol $B_0(t)$,
\begin{equation} 
\label{eq:transportequation}
\frac{\partial}{\partial t} B_0(t) - \{ H_0,B_0(t) \} 
- \ui\,[H_1,B_0(t)] = 0 \ ,
\end{equation} 
which for Hamiltonians with a scalar sub-principal symbol $H_1$ reduces to 
\begin{equation} 
\frac{\partial}{\partial t} B_0(t) - \{ H_0,B_0(t) \} = \frac{\ud}{\ud t}
\,( B_0(t)\circ\Phi^{-t} ) = 0\ .
\end{equation} 
In order to solve the general case, (\ref{eq:transportequation}) is 
rewritten as
\begin{equation} 
\label{altern_dgl}
\frac{\ud}{\ud t}\,\left[ d^{-1}(\vecp,\vecx,-t)\,
B_0(t)(\Phi^{-t}(\vecp,\vecx))\,d(\vecp,\vecx,-t) \right] = 0 \ ,
\end{equation}
where $d$ has to fulfill 
\begin{equation}
\label{spintrans} 
\dot{d}(\vecp,\vecx,t) + \ui\,H_1(\Phi^t(\vecp,\vecx))\, 
d(\vecp,\vecx,t) = 0\ , \qquad d(\vecp,\vecx,0) = \eins_n \ .
\end{equation}
Here the time derivative has to be understood along the trajectory 
$\Phi^t(\vecp,\vecx)$. The quantity $d$ already appeared in 
\cite{BolKep99a,BolKep99b}, where it was introduced in order
to describe the (semiclassical) propagation of the spin degrees of
freedom along the trajectories of the Hamiltonian flow $\Phi^t$.

We are now in a position to calculate the principal symbol of 
$\hat{B}(t)$ from (\ref{altern_dgl}),
\begin{equation} 
B_0(t)(\vecp,\vecx) = d(\Phi^t(\vecp,\vecx),-t)\,B_0(\Phi^t(\vecp,\vecx))\,
d^{-1}(\Phi^t(\vecp,\vecx),-t) \ .
\end{equation}
Equivalently, employing the property
\begin{equation}
d(\Phi^t(\vecp,\vecx),-t) = d^{-1}(\vecp,\vecx,t) 
\end{equation}
that can be deduced from (\ref{spintrans}) (see, e.g., \cite{BruNou99})
one obtains
\begin{equation} 
\label{eq:principal_symbol}
B_0(t)(\vecp,\vecx) = d^{-1}(\vecp,\vecx,t)\,B_0(\Phi^t(\vecp,\vecx))\, 
d(\vecp,\vecx,t)\ .
\end{equation}
Notice that in \cite{BruNou99} the quantity $\Ga(t,\vecp,\vecx)
=d^{-1}(\vecp,\vecx,t)$ is used instead of $d(\vecp,\vecx,t)$. If we take 
into account that $d(\vecp,\vecx,t)$ is unitary, which follows from 
(\ref{spintrans}) since $H_1$ is hermitian (see, e.g.,
\cite{BolKep99a,BruNou99}), then (\ref{eq:principal_symbol}) 
also reads
\begin{equation}
\label{Egorlead}
B_0(t)(\vecp,\vecx) = d^\dagger(\vecp,\vecx,t)\,B_0(\Phi^t(\vecp,\vecx))\, 
d(\vecp,\vecx,t)\ ,
\end{equation}
which is the form that we will use below.

In principle one could determine all coefficients $B_l(t)$ in this fashion, 
but one must be aware of the fact that the resulting symbol $B(t)$ will in 
general not be in $\scl^k(1)$, unless one restricts the growth of the symbol
$H$ in such a way that one would exclude the Pauli Hamiltonians 
(\ref{Pauliop}) that we are interested in. If, however, one restricts 
attention to observables whose symbols have compact support, 
$B\in C_0^\infty (\rz^d\times\rz^d)\otimes\kz^{n\times n}$, this problem is
avoided and one obtains a Weyl symbol $B_{sum}(t)\in C_0^\infty 
(\rz^d\times\rz^d)\otimes\kz^{n\times n}$ by, say, Borel summation of the 
asymptotic expansion (\ref{Boftasym}), with the solutions $B_l(t)$ of the 
recursive Cauchy problem (\ref{eq:cauchy_problems}) entering. On the level 
of the corresponding operators one therefore finds
\begin{equation} 
\left\|  \widehat{B_{sum}(t)} - \ue^{\frac{\ui}{\hbar}\hat{H}t}\,\hat{B}\,
\ue^{-\frac{\ui}{\hbar}\hat{H}t} \right\|_{\cL} \le C\,\hbar^N\ , \qquad
\text{for all } N\in\nz \ ,
\end{equation}
where $\|\cdot\|_\cL$ denotes the norm on $\cL(L^2(\rz^d)\otimes\kz^n, 
L^2(\rz^d)\otimes\kz^n)$.
  
The same result, which is a variant of the Egorov Theorem \cite{Ego69},  
can be achieved for general $B\in\scl^k(1)$ under suitable requirements on 
the symbol of the Hamiltonian $\hat{H}$ (see \cite[ch.2.3]{Ivr98}):
\begin{prop}
\label{Egorov} 
Let $H$ be in $\skl(m)$ and $B\in\scl^k(1)$ and suppose that
\begin{equation} 
\label{eq:assumption_H}
\left\| \partial_p^\al\partial_x^\be H_j(\vecp,\vecx) \right\| 
\le C \quad\text{for all}\quad (\vecp,\vecx)\in\rz^d\times\rz^d \quad 
\text{and}\quad |\al| + |\be| + 2j \ge 2 \ .
\end{equation}
Then the estimate 
\begin{equation} 
\left\| \widehat{B_{sum}(t)} - \ue^{\frac{\ui}{\hbar}\hat{H}t}\,\hat{B}\, 
\ue^{-\frac{\ui}{\hbar}\hat{H}t} \right\|_\cL \le D\,\hbar^N
\end{equation}
holds for arbitrary $N>0$ and $t\in [0,T]$. 
\end{prop}
Let us remark that under the assumption (\ref{eq:assumption_H}) the 
Hamiltonian vector field $X_{H_{0,s}}$ grows at most linearly at infinity.
Therefore, a trajectory $\Phi^t(\vecp,\vecx)$ cannot blow up at finite
times so that the flow exists globally on $\rz^d\times\rz^d$ (see, e.g., 
\cite{Rob87}). We have not made any attempt at improving the bounds
on the time $T$, in terms of $\hbar$, up to which Proposition \ref{Egorov} 
holds, since for our further purposes we are mainly interested in the relation
(\ref{Egorlead}). For the scalar case such improvements have, however, been 
established in \cite{BamGraPau99,BouRob99}.

As a second essential input for the proof of quantum ergodicity, in addition 
to the Egorov property (\ref{Egorlead}), we require a Szeg\"o limit formula 
(see, e.g., \cite{Gui79}) that connects averaged expectation values of an 
observable semiclassically with a classical average. On the quantum mechanical 
side we consider an interval $I(E,\hbar):=[E-\hbar\om,E+\hbar\om]$, with some
$\om >0$, such that $I(E,\hbar)\subseteq [E-\ve,E+\ve]$ if $\hbar$ is 
sufficiently small. According to the assumptions 1.--4. on the Hamiltonian
$\hat H$ in section \ref{sec2} the spectrum of $\hat H$ in $I(E,\hbar)$
is therefore discrete. We then denote the number of eigenvalues contained 
in $I(E,\hbar)$ by $N_I$, and $\{\psi_k\}$ shall be the (orthonormal) 
eigenvectors of $\hat H$ associated with the eigenvalues $E_k\in I(E,\hbar)$.
On the classical side, let $\Om_E := H_{0,s}^{-1}(E)$ be the hypersurface 
of energy $E$ in phase space and denote the normalised Liouville measure 
on $\Om_E$ by $\ud\mu_E$, i.e.
\begin{equation}
\ud\mu_E (\vecp,\vecx) = \frac{1}{\vol\Om_E}\,\de\bigl( H_{0,s}(\vecp,\vecx)
-E \bigr)\ \ud p\,\ud x\ .
\end{equation}
In the following we will abbreviate averages of (smooth) matrix valued
functions $B\in C^\infty (\rz^d\times\rz^d)\otimes\kz^{n\times n}$ over
$\Om_E$ by
\begin{equation}
\mu_E (B) := \int_{\Om_E} B(\vecp,\vecx)\ \ud\mu_E (\vecp,\vecx) \ .
\end{equation}
From now on we also suppose that the Hamiltonian flow $\Phi^t$ generated
by $H_{0,s}$ is ergodic with respect to $\mu_E$, i.e.\ for $f\in 
L^1(\Om_E,d\mu_E)$ and $\mu_E$-almost all $(\vecp,\vecx)\in\Om_E$
\begin{equation}
\label{Phiergodic}
\lim_{T\to\infty}\frac{1}{T}\int_0^T f\bigl(\Phi^t (\vecp,\vecx)\bigr)\ \ud t
= \int_{\Om_E} f(\vecq,\vecy)\ \ud\mu_E (\vecq,\vecy) \ . 
\end{equation}
This in particular implies that the set of periodic points of $\Phi^t$ 
with periods $T>0$ has Liouville measure zero.
 
The rest of this section will now be devoted to the proof of
\begin{prop}[Szeg\"o limit formula]
\label{Szegoe}
Let $\hat H$ be a quantum Hamiltonian with symbol $H\in\scl^0(m)$, where
$m\geq 1$, that fulfills the properties 1.--4. of section \ref{sec2}. If
then $\hat B$ is an observable with symbol $B\in\scl^0(1)$ and principal
symbol $B_0$, the following Szeg\"o limit formula holds,
\begin{equation}
\label{Szego}
\lim_{\hbar\to 0}\,\frac{1}{N_I}\sum_{E_k\in I(E,\hbar)} \langle\psi_k,
\hat B\psi_k\rangle = \frac{1}{n}\,\mtr\mu_E (B_0) \ .  
\end{equation}
\end{prop}
\begin{proof}
In principle we adopt the method presented in \cite{DimSjo99}, and modify 
it appropriately where necessary. Let us first recall that if $g\in 
C_0^\infty (\rz)$ is suitably chosen, with $g(\la)=\la$ on a neighbourhood 
of the interval $I(E,\hbar)$, the operator $g(\hat H)$ has the same 
spectrum in $I(E,\hbar)$ as $\hat H$ itself. Furthermore, the symbol 
$H_g\in\scl^0 (1)$ of $g(\hat H)$ has an asymptotic expansion that coincides 
on $H_{0,s}^{-1}(I(E,\hbar))$ with that of $H$. Since below we are
localising in energy to the interval $I(E,\hbar)$ so that we can consider
$g(\hat H)$ instead of $\hat H$, from now on we simply suppose that 
$H\in\scl^0 (1)$.

Let now $\chi\in C_0^\infty (\rz)$ be given with $\chi\equiv 1$ on 
$I(E,\hbar)$ and such that the spectrum of $\hat H$ in supp $\chi$ is 
discrete. This might require to choose $\hbar$ small enough. Then consider 
the (energy localised) quantum mechanical time evolution operator
\begin{equation}
\hat U_\chi (t) := \ue^{-\frac{\ui}{\hbar}\hat H t}\,\chi(\hat H) \ .
\end{equation}
Up to an error of order $O(\hbar^\infty)$ in trace norm, this operator
can be approximated by a semiclassical Fourier integral operator with
kernel
\begin{equation}
\label{Fourior}
K_\chi (\vecx,\vecy,t) = \frac{1}{(2\pi\hbar)^d}\int_{\rz^d} a_\hbar
(\vecx,\vecy,t,\vecxi)\,\ue^{\frac{\ui}{\hbar}(S(\vecx,\vecxi,t)-
\vecxi\cdot\vecy)}\ \ud\xi \ ,
\end{equation}
if $|t|$ is small enough (see, e.g., \cite{DimSjo99}). We remark that here 
the amplitude $a_\hbar$ takes values in $\kz^{n\times n}$, thus it also 
represents the internal (i.e.\ spin) degrees of freedom. As explained in 
\cite{BolKep99a}, the phase $S$ in (\ref{Fourior}) is then given as the 
solution of the Hamilton-Jacobi equation
\begin{equation}
H_{0,s}\bigl(\partial_x S(\vecx,\vecxi,t),\vecx\bigr) + \partial_t S
(\vecx,\vecxi,t) = 0 \ ,\qquad S(\vecx,\vecxi,0) = \vecx\cdot\vecxi \ .
\end{equation}
In leading semiclassical order the transport equation for $a_\hbar\sim
a_0 +\hbar\,a_1 +\dots$ can be solved by a separation of the internal
degrees of freedom from the translational ones. The latter lead to the
expression known from the scalar case (see, e.g., \cite{DimSjo99}),
whereas the modifications required by the matrix character of $a_0$
are provided by the solution $d(\vecp,\vecx,t)$ of (\ref{spintrans}),
see \cite{BolKep99a}. For the present purpose, however, one only needs the
initial condition $a_\hbar|_{t=0}=\chi(H_{0,s})\eins_n+O(\hbar)$.

In a next step we consider $\rho\in C^\infty(\rz)$ with Fourier transform
$\tilde\rho\in C_0^\infty(\rz)$ such that
\begin{equation}
\tr\frac{1}{2\pi}\int_\rz \tilde\rho(t)\,\ue^{\frac{\ui}{\hbar}Et}\,\hat B
\,\hat U_\chi(t)\ \ud t = \sum_k\chi(E_k)\,\langle\psi_k,\hat B\psi_k\rangle
\,\rho\left(\frac{E_k -E}{\hbar}\right) \ ,
\end{equation}
where $\tr\,(\cdot)$ denotes the operator trace on $L^2(\rz^d)\otimes\kz^n$.
Approximating $\hat U_\chi(t)$ with the help of (\ref{Fourior}), in leading
semiclassical order one then has to calculate
\begin{equation}
\label{scInt}
\frac{1}{2\pi\,(2\pi\hbar)^d}\int_{\rz}\int_{\rz^d}\int_{\rz^d}
\tilde\rho(t)\,\mtr\bigl(B_0(\partial_x S,\vecx)\,a_0 (\vecx,\vecx,t,\vecxi)
\bigr)\,\ue^{\frac{\ui}{\hbar}(S(\vecx,\vecxi,t)-\vecxi\cdot\vecx+Et)}\ 
\ud\xi\,\ud x\,\ud t
\end{equation}
with the method of stationary phase. The stationary points of the phase
$S(\vecx,\vecxi,t)-\vecxi\cdot\vecx+Et$ are given by $(\vecxi_{st},\vecx_{st},
t_{st})\in\rz^d\times\rz^d\times\rz$ such that $(\vecxi_{st},\vecx_{st})
\in\Om_E$ is a periodic point of the Hamiltonian flow $\Phi^t$ with
period $t_{st}$. Since we have assumed that $E$ is not a critical value
for $H_{0,s}$, the periods of $\Phi^t$ on $\Om_E$ do not accumulate at
zero, see \cite{Rob87}. Hence, if the support of $\tilde\rho$ is chosen 
small enough, the
manifold of critical points contributing to (\ref{scInt}) is given by
$\Om_E\times\{0\}$. In analogy to \cite{BolKep99a} we thus obtain
\begin{equation}
\label{scWeyl}
\sum_k\chi(E_k)\,\langle\psi_k,\hat B\psi_k\rangle\,\rho\left(
\frac{E_k -E}{\hbar}\right) = \chi(E)\,\frac{\tilde\rho(0)}{2\pi}
\frac{\vol\Om_E}{(2\pi\hbar)^{d-1}}\,\bigl( \mtr\mu_E(B_0) 
+ O(\hbar) \bigr) \ .  
\end{equation}
Moreover, since we require $H_{0,s}$ to be such that no $E'\in [E-\ve,E+\ve]$ 
is a critical value and all $\Om_{E'}$ are compact, (\ref{scWeyl}) holds 
true with $E$ replaced by $E'$ uniformly in $[E-\ve,E+\ve]$.  
In a last step we now apply the Tauberian Lemma of \cite{BruPauUri95},
which takes non-negative weights into account. To this end we have to 
ensure that $\langle\psi_k,\hat B\psi_k\rangle$ is non-negative. However,
since $\hat B$ is bounded, this can always be achieved by adding a suitable 
constant. Moreover, according to our above choice of $\chi$ we find that 
$\chi(E)=1$ and $\chi(E_k)=1$ for all $E_k\in I(E,\hbar)$. Therefore
\begin{equation}
\label{Matrixasy}
\sum_{E_k\in I(E,\hbar)}\langle\psi_k,\hat B\psi_k\rangle = 
\frac{\om}{\pi}\,\frac{\vol\Om_E}{(2\pi\hbar)^{d-1}}\,\mtr\mu_E(B_0) 
+ o(\hbar^{1-d})  \ .
\end{equation}
Repeating the above reasoning with the identity instead of $\hat B$,
one can express the number $N_I$ of eigenvalues in $I(E,\hbar)$ 
semiclassically, 
\begin{equation}
\label{NIasy}
N_I = \frac{n\om}{\pi}\,\frac{\vol\Om_E}{(2\pi\hbar)^{d-1}}\,
+ o(\hbar^{1-d})  \ .
\end{equation}
Thus, (\ref{Matrixasy}) and (\ref{NIasy}) together finally yield the 
Szeg\"o limit formula (\ref{Szego}).
\end{proof}

\section{Quantum ergodicity}
\label{sec4}

In the following we will restrict attention to the case of spin $s=1/2$, 
which is both the simplest and physically most important situation. This
means that below $n=2s+1=2$ will be chosen, so that all symbols of
observables and Hamiltonians take values in the hermitian $2\times 2$ 
matrices. We also assume that the quantum Hamiltonian describes the
coupling of translational and spin degrees of freedom as in (\ref{Pauliop}).
This restricts the subprincipal symbol $H_1$ of $\hat H$ to be a traceless 
hermitian matrix. As a consequence, the spin transport equation 
(\ref{spintrans}) is solved by a spin transport matrix $d(\vecp,\vecx,t)
\in\SU (2)$.

Our strategy of approaching quantum ergodicity is inspired by the method 
introduced in \cite{Zel96,ZelZwo96}, which does not require to rely
on a positive quantisation, such as anti-Wick or Friedrichs quantisation.
It is rather based on an analysis of the expression
\begin{equation} 
\label{eq:S2_defn} 
S_2(E,\hbar):= \frac{1}{N_I} \sum_{E_k \in I(E,\hbar)} 
\Bigl| \langle \psi_k, \hat{B} \psi_k \rangle - \frac{1}{2} \mtr \mu_E(B_0) 
\Bigr|^2 \ ,
\end{equation}
which is the variance of the expectation values of the quantum observable 
$\hat{B}$ about the classical mean value of its principal symbol $B_0$.
We are in particular interested in the behaviour of (\ref{eq:S2_defn}) in 
the limit $\hbar\to 0$. For this purpose we introduce the bounded and
self-adjoint auxiliary operator 
\begin{equation} 
\label{eq:aux_op}
\hat{B}_T := \frac{1}{T} \int_0^T \hat{U}^\dagger(t)\,\hat{B}\,\hat{U}(t) 
\ \ud t - \frac{1}{2} \mtr \mu_E(B_0)\,\eins_2 \ .
\end{equation} 
Its expectation values in eigenstates of the Hamiltonian are
\begin{equation}
\label{expectaux} 
\langle \psi_k, \hat{B}_T \psi_k \rangle = \langle \psi_k, \hat{B} 
\psi_k \rangle - \frac{1}{2} \mtr \mu_E (B_0) \ ,
\end{equation}
such that
\begin{equation}
\label{S2withBT}
S_2(E,\hbar) = \frac{1}{N_I} \sum_{E_k \in I(E,\hbar)} \Bigl| \langle \psi_k,
\hat{B}_T \psi_k \rangle \Bigr|^2 \ .
\end{equation} 
Using the Egorov property (\ref{Egorlead}), the principal symbol of the 
auxiliary operator reads
\begin{equation}
\label{eq:aux_op_ps}
B_{T,0}(\vecp,\vecx) = \frac{1}{T} \int_0^T d^\dagger(\vecp,\vecx,t)\, 
B_0(\Phi^t(\vecp,\vecx))\,d(\vecp,\vecx,t) \ \ud t - 
\frac{1}{2} \mtr \mu_E (B_0)\,\eins_2 \ .
\end{equation} 
Comparing this expression with the analogous one obtained in the scalar 
case, one observes that the principal symbol $B_0$ is not only transported
by the flow $\Phi^t$, but also conjugated with the spin transport matrix 
$d(\vecp,\vecx,t)$. Thus both the translational and the spin dynamics are 
involved. This observation suggests that the ergodicity of the flow 
$\Phi^t$ will no longer suffice to yield quantum ergodicity. One therefore
has to combine the classical dynamics of the translational degrees of 
freedom and the spin dynamics in a suitable way, and then one demands ergodic 
properties of the combined dynamics. In order to achieve this we employ the 
same construction as in \cite{BolKep99b} and hence introduce the product 
phase space 
\begin{equation}
\cM:= \Om_E \times \SU(2) \ .
\end{equation}
The combined flow $Y^t\,:\cM\to\cM$ is then defined as an $\SU(2)$-extension 
of the Hamiltonian flow $\Phi^t$ on $\Om_E$, i.e.\ for $(\vecp,\vecx)\in 
\Om_E$ and $g\in\SU(2)$ we set
\begin{equation} 
Y^t((\vecp,\vecx),g):=\bigl(\Phi^t(\vecp,\vecx), d(\vecp,\vecx,t)g\bigr) \ .
\end{equation}
The initial condition $Y^0 =\mbox{id}$ is obviously fulfilled, and $Y^{t+s}
=Y^t\circ Y^s$ follows from the composition law
\begin{equation} 
d(\vecp,\vecx,t+s) = d\bigl(\Phi^t(\vecp,\vecx),s\bigr)\,d(\vecp,\vecx,t)
\end{equation}
that derives from the spin transport equation (\ref{spintrans}). In ergodic 
theory such a combined dynamics is also known as a skew product (see, e.g., 
\cite{CorFomSin82}). On $\cM$ the product measure $\mu:=\mu_E\times\mu_H$ 
consisting of the Liouville measure $\mu_E$ on $\Om_E$ and the normalised 
Haar measure $\mu_H$ on $\SU(2)$ is introduced. This measure is normalised 
and invariant under the flow $Y^t$, since $\mu_E$ is normalised and invariant 
under $\Phi^t$ and $\mu_H$ is both left and right invariant. Ergodicity of 
$Y^t$ on $\cM$ with respect to $\mu$ then means that for 
$F\in L^1(\cM,\ud\mu)$ 
\begin{equation}
\label{Yergodic}
\lim_{T\to\infty} \frac{1}{T} \int_0^T F\bigl(Y^t((\vecp,\vecx),g)\bigr) 
\ \ud t = \int_\cM F((\vecq,\vecy),h) \ \ud\mu((\vecq,\vecy),h)
\end{equation}
holds for $\mu$-almost all $((\vecp,\vecx),g)\in\cM$. In particular, if
one chooses $F$ to be independent of $g$, then (\ref{Yergodic}) reduces
to the condition (\ref{Phiergodic}), so that the ergodicity of the extended 
flow $Y^t$ on $\cM$ implies the ergodicity of the flow $\Phi^t$ on the
base manifold $\Om_E$.  

Our main result now states the effect of the ergodicity of $Y^t$ on the
semiclassical asymptotics of eigenfunctions of $\hat{H}$.
\begin{theorem}[Quantum ergodicity] 
\label{QE} 
Let $\hat{H}$ be a quantum Hamiltonian with symbol $H\in\scl^0(m)$, where 
$m\ge 1$, and principal symbol $H_0=H_{0,s}\eins_2$, which fulfills the 
conditions 1.--4.\ of section \ref{sec2} and, furthermore, satisfies
\begin{equation} 
\tag*{(\ref{eq:assumption_H})}
\| \partial_p^\al \partial_x^\be H_j(\vecp,\vecx) \| \le C
\quad\text{for all}\quad (\vecp,\vecx) \in \rz^d \times \rz^d \quad 
\text{and} \quad |\al|+|\be|+ 2j \ge 2.
\end{equation}
Then, under the condition that $Y^t$ is ergodic on $\cM$ with respect to 
$\mu$, in every sequence $\{\psi_k \mid E_k \in I(E,\hbar)\}$ of orthonormal 
eigenfunctions of $\hat{H}$ there exists a subsequence $\{\psi_{k_j} \mid
E_{k_j} \in I(E,\hbar) \}$ of density one, i.e.
\begin{equation} 
\lim_{\hbar\to 0} \frac{\# \{j \mid E_{k_j} \in I(E,\hbar) \}}
{\# \{k \mid E_k \in I(E,\hbar) \}} = 1\ ,
\end{equation}
such that for every quantum observable $\hat B$ with hermitian symbol 
$B\in\scl^0(1)$ and principal symbol $B_0$ 
\begin{equation} 
\label{qe}
\lim_{j\to\infty}\ \langle \psi_{k_j}, \hat{B} \psi_{k_j} \rangle 
= \frac{1}{2} \mtr \mu_E(B_0)\ .
\end{equation}
Moreover, the subsequence $\{\psi_{k_j}\}$ can be chosen independent of 
the observable $\hat B$.
\end{theorem}
\begin{proof} 
An application of the Cauchy-Schwartz inequality on the right-hand side
of (\ref{S2withBT}) yields an upper bound for the quantity (\ref{eq:S2_defn}) 
that reads 
\begin{equation}
\label{S2bound} 
S_2(E,\hbar) \le \frac{1}{N_I} \sum_{E_k\in I(E,\hbar)} 
\langle \psi_k,(\hat{B}_T)^2 \psi_k \rangle \ .
\end{equation}
In order to determine the limit as $\hbar\to 0$ of (\ref{S2bound})
one can now apply Proposition \ref{Szegoe}, which gives 
\begin{equation}
\label{trsquare}
\lim_{\hbar\to 0} S_2(E,\hbar) \le \frac{1}{2} \mtr\mu_E
\bigl((B_{T,0})^2\bigr)\ .
\end{equation}
Quantum ergodicity then follows, if the bound on the right-hand side can
be shown to vanish. This will indeed be possible in the limit $T\to\infty$.
In order to achieve this we first employ the ergodicity of $Y^t$ in that 
we choose $F((\vecq,\vecy),h)=h^{\dagger}\,B_0(\vecq,\vecy)\,h$ in the 
relation (\ref{Yergodic}), and thus find
\begin{equation}
\label{Yergcon}
\lim_{T\to\infty} \frac{1}{T} \int_0^T g^{\dagger}\,d^{\dagger}
(\vecp,\vecx,t)\, 
B_0(\Phi^t(\vecp,\vecx))\,d(\vecp,\vecx,t)\,g \ \ud t = 
\int_{\SU(2)} h^{\dagger}\,\mu_E(B_0)\,h \ \ud\mu_H(h) 
\end{equation} 
for $\mu_E$-almost all $(\vecp,\vecx)\in\Om_E$ and $\mu_H$-almost all
$g\in\SU(2)$. In terms of the principal symbol (\ref{eq:aux_op_ps}) of the 
auxiliary operator $\hat{B}_T$ this means
\begin{equation}
\label{princBTlim}
\lim_{T\to\infty} g^\dagger\,B_{T,0}(\vecp,\vecx)\,g = \int_{\SU(2)} 
h^{\dagger}\,\mu_E(B_0)\,h \ \ud\mu_H(h) - \frac{1}{2} \mtr \mu_E(B_0) \, 
\eins_2 \ .
\end{equation} 

Our next goal is to calculate the right-hand side of (\ref{princBTlim}). 
To this end we represent the hermitian $2\times 2$ matrix $\mu_E(B_0)$ 
as a linear combination of $\eins_2$ and the Pauli matrices $\si_k$, i.e.\ 
$\mu_E(B_0) = \frac{1}{2}\mtr\mu_E(B_0) + \vecb\cdot\vecsig$, where 
$\vecb\in\rz^3$. We then recall that for every $g\in\SU(2)$ the adjoint map
$\Ad_g\,:\su(2)\to\su(2)$ is defined as $\Ad_g(X)=g^{\dagger}\,X\,g$. Thus,
$\Ad_h(\vecb\cdot\vecsig)\in\su(2)$ can be expanded in terms of the Pauli
matrices $\si_k$ such that $h^{\dagger}\,\vecb\cdot\vecsig\,h = (\vp(h)\vecb)
\cdot\vecsig$. The map $\vp:\,\SU(2)\to\SO(3)$ that results in this way
can be identified as the universal (twofold) covering of $\SO(3)$ by 
$\SU(2)$. Therefore
\begin{equation}
\label{detBTrep}
\int_{\SU(2)} h^{\dagger}\,\mu_E(B_0)\,h \ \ud\mu_H(h) = 
\frac{1}{2}\mtr\mu_E(B_0)\,\eins_2 + \left( \int_{\SU(2)}\vp(h)\ \ud\mu_H(h) 
\,\vecb \right)\cdot \vecsig \ .  
\end{equation} 
We now show that the second term on the right-hand side of (\ref{detBTrep})
vanishes. To this end we multiply the integral over $\SU(2)$, which
yields a $3\times 3$ matrix, with an arbitrary orthogonal matrix $R\in\SO(3)$ 
from the left. We then exploit the fact that there exists $\tilde g\in\SU(2)$ 
such that $R=\vp(\tilde g)$, together with the left invariance of the Haar 
measure, in order to conclude that
\begin{equation}
R\int_{\SU(2)}\vp(h)\ \ud\mu_H(h) = \int_{\SU(2)}\vp(\tilde g h)\ \ud\mu_H(h)
=\int_{\SU(2)}\vp(h)\ \ud\mu_H(h)\ .
\end{equation} 
Since hence the $3\times 3$ matrix represented by the integral over 
$\SU(2)$ is invariant under left multiplication by an arbitrary element 
of $\SO(3)$, it must be the zero matrix. Therefore, the right-hand side of
(\ref{princBTlim}) vanishes.

We now choose some $g\in\SU(2)$ such that (\ref{Yergcon}) holds and therefore
obtain that
\begin{equation}
\lim_{T\to\infty} \mtr \bigl(B_{T,0}(\vecp,\vecx)\bigr)^2 = 
\lim_{T\to\infty} \mtr \bigl( g^\dagger\,B_{T,0}(\vecp,\vecx)\,g \bigr)^2 
= 0 
\end{equation}
holds for $\mu_E$-almost all $(\vecp,\vecx)\in\Om_E$. Thus, after an 
integration over $\Om_E$, this together with (\ref{trsquare}) implies that 
\begin{equation}
\lim_{\hbar\to 0} S_2(E,\hbar) = 0\ . 
\end{equation}
According to a standard argument in the proof of quantum ergodicity for
scalar Hamiltonians, see \cite{Zel87,Col85}, the vanishing of $S_2(E,\hbar)$
in the semiclassical limit implies the existence of a density-one
subsequence $\{\psi_{k_j}\}_{j\in\nz}\subset\{\psi_k\}_{k\in\nz}$
such that (\ref{qe}) holds. Another standard, diagonal construction
then ensures that a subsequence can be chosen that is independent of the
observable, see \cite{Zel87,Col85}.
\end{proof} 

\section{Discussion}
\label{sec5}

In the case of scalar Hamiltonians quantum ergodicity is often interpreted
in terms of Wigner- or Husimitransforms of eigenfunctions. In this
context one concludes that along a subsequence of density one the
Wigner- or Husimitransforms of eigenfunctions of a quantum ergodic
Hamiltonian weakly converge, as distributions or measures, respectively, 
to Liouville measure. Thus the lifts of eigenfunctions to phase space
become equidistributed on the hypersurface $\Om_E$ of energy $E$. In the
present situation an analogous interpretation first requires to introduce
matrix valued Wigner- and Husimitransforms. In general, the Wignertransform
of $\psi\in\cS'(\rz^d)\otimes\kz^n$ is given by
\begin{equation}
W[\psi](\vecp,\vecx) = \int_{\rz^d}\ue^{-\frac{\ui}{\hbar}\vecp\cdot\vecy}
\,\overline{\psi}\bigl( \vecx-\frac{1}{2}\vecy \bigr)\otimes\psi
\bigl( \vecx+\frac{1}{2}\vecy \bigr) \ \ud y \ .
\end{equation}
Then expectation values of observables $\hat B$ with symbols $B\in\skl^k(1)$
in states described by $\psi\in L^2(\rz^d)\otimes\kz^n$ read
\begin{equation}
\label{ExpectWig}
\langle \psi,\hat B\psi \rangle = \frac{1}{(2\pi\hbar)^d}\int_{\rz^d}
\int_{\rz^d}\mtr\bigl( W[\psi](\vecp,\vecx)\,B(\vecp,\vecx) \bigr)\ 
\ud p\,\ud x \ . 
\end{equation}
In order to convert the statement of Theorem \ref{QE} into one about 
the matrix components of Wignertransforms we introduce the special observables 
$\hat{B}^{(rs)}$ with symbols $B^{(rs)}=b\,E_{rs}\in\skl^0(1)$, where $b$ is 
a real valued function on phase space that is independent of $\hbar$, and 
the four constant matrices $E_{rs}$ are defined by 
\begin{equation}
E_{11} = \begin{pmatrix} 1&0\\0&0 \end{pmatrix}\ ,\quad
E_{12} = \begin{pmatrix} 0&1\\0&0 \end{pmatrix}\ ,\quad
E_{21} = \begin{pmatrix} 0&0\\1&0 \end{pmatrix}\ ,\quad
E_{22} = \begin{pmatrix} 0&0\\0&1 \end{pmatrix}\ .
\end{equation}
Although the off-diagonal symbols $b\,E_{12}$ and $b\,E_{21}$ are
non-hermitian, Theorem \ref{QE} can be applied to all of the observables 
$\hat{B}^{(rs)}$ in an obvious manner. Together with the relation 
(\ref{ExpectWig}) this then reveals that along the subsequence 
$\{\psi_{k_j}\}$ of density one specified in the Theorem the matrix
components of the Wignertransforms of the eigenfunctions weakly converge, 
as distributions on $C^\infty_0(\rz^d\times\rz^d)$, to either Liouville 
measure or to zero. More specifically,
\begin{equation}
\label{Wigsc}
\lim_{j\to\infty}\,\frac{1}{(2\pi\hbar)^d}W[\psi_{k_j}] = \frac{1}{2}
\frac{1}{\vol\Om_E}\,\de\bigl( H_{0,s}-E \bigr)\,\eins_2 
\end{equation}
component-wise in $\cD'(\rz^d\times\rz^d)$.
In particular, the semiclassical limit of the Wignerdistributions is a 
scalar multiple of the identity matrix. This means that the `spin up'
and `spin down' components become identical and equidistributed over
the hypersurface of energy $E$ in phase space. Moreover, there occurs no 
mixture between `spin up' and `spin down' components, which can also be 
seen on the right-hand side of (\ref{qe}) since there only the diagonal 
elements of the principal symbol $B_0$ contribute.

Upon introducing matrix valued Husimitransforms through
\begin{equation}
H[\psi](\vecp,\vecx) = \frac{1}{(\pi\hbar)^d}\int_{\rz^d}\int_{\rz^d} 
W[\psi](\vecq,\vecy)\,\ue^{-\frac{1}{\hbar}((\vecp-\vecq)^2+(\vecx-\vecy)^2)}
\ \ud q\,\ud y\ ,
\end{equation}
one can also consider anti-Wick quantisations of symbols $B\in\skl^0(1)$. 
Expectation values of the corresponding anti-Wick operators $\hat{B}_{AW}$
then read in analogy to (\ref{ExpectWig})
\begin{equation}
\label{ExpectHus}
\langle \psi,\hat{B}_{AW}\psi \rangle = \frac{1}{(2\pi\hbar)^d}\int_{\rz^d}
\int_{\rz^d}\mtr\bigl( H[\psi](\vecp,\vecx)\,B(\vecp,\vecx) \bigr)\ 
\ud p\,\ud x \ . 
\end{equation}
Since, moreover, for $B\in\skl^0(1)$ 
\begin{equation}
\left\| \hat B - \hat{B}_{AW} \right\|_{\cL} = O(\hbar) \ ,
\end{equation}
one can replace $\hat B$ by $\hat{B}_{AW}$ in (\ref{qe}) so that 
Theorem~\ref{QE} implies an analogue of (\ref{Wigsc}), i.e.%
\begin{equation}
\label{Hussc}
\lim_{j\to\infty}\,\frac{1}{(2\pi\hbar)^d}H[\psi_{k_j}]\ \ud p\,\ud x = 
\frac{1}{2}\,\eins_2\,\ud\mu_E \ ,
\end{equation}
where the convergence has to be understood component-wise as a weak 
convergence of probability measures on the phase space $\rz^d\times\rz^d$.  

As a further point we now want to discuss whether it is actually necessary
to introduce the $\SU(2)$-extension $Y^t$ of the Hamiltonian flow $\Phi^t$,
and to require ergodicity of $Y^t$ in order to obtain quantum ergodicity.
To this end we provide an example showing that in general it would not 
suffice to demand only ergodicity of $\Phi^t$. Let us therefore introduce 
a quantum Hamiltonian $\hat H$ of the type introduced in sections 
\ref{sec2}.~-- \ref{sec4}.\ with symbol\footnote{We owe this example to 
Stefan Keppeler.}
\begin{equation}
H(\vecp,\vecx) = H_{0,s}(\vecp,\vecx)\,\eins_2 + 
\hbar\,C(\vecp,\vecx)\,\si_j \ ,
\end{equation}
where $\si_j$ is one of the Pauli matrices and $H_{0,s}$ shall be chosen such
that $\Phi^t$ is ergodic on $\Om_E$. Now $\si_j$ can obviously also be 
considered as a bounded self-adjoint operator on $L^2(\rz^d)\otimes\kz^2$
which commutes with $\hat H$; in fact, $\frac{\hbar}{2}\si_j$ is the $j$-th
component of the spin observable. Hence, one can introduce joint eigenvectors
$\psi_k\in L^2(\rz^d)\otimes\kz^2$ of $\hat H$ and $\si_j$. Since $\si_j^2
=\eins_2$, the eigenvalues of $\si_j$ are $\la_k=\pm 1$. Introducing $U_j
\in\U(2)$ such that $U_j\si_jU_j^\dagger$ is diagonal, we can switch to
eigenvectors $\vp_k := U_j\psi_k$ such that 
\begin{equation}
\vp_k = \begin{pmatrix} \vp_k^{(+)} \\ 0 \end{pmatrix} \quad \text{if}
\quad \la_k = +1 \qquad \text{and} \qquad 
\vp_k = \begin{pmatrix} 0 \\ \vp_k^{(-)} \end{pmatrix} \quad \text{if}
\quad \la_k = -1 \ .
\end{equation}
The expectation values of an observable $\hat B$ in these eigenvectors
$\vp_k$ therefore read
\begin{equation}
\langle \vp_k,\hat B\vp_k \rangle = 
\begin{cases} ( \vp_k^{(+)},\hat{B}_{11}\vp_k^{(+)} ) &
\text{if $\la_k =+1$}\ , \\
( \vp_k^{(-)},\hat{B}_{22}\vp_k^{(-)} ) &
\text{if $\la_k =-1$}\ , \end{cases}
\end{equation}
where $(\cdot,\cdot)$ denotes the scalar product in $L^2(\rz^d)$. Since
the subsequences of the $\vp_k$'s with $\la_k=+1$ and $\la_k=-1$,
respectively, are each of density one half, quantum ergodicity cannot
hold for general observables $\hat B$. 

On the other hand, the above example does not fulfill the requirements of
Theorem \ref{QE} so that no contradiction occurs. To see this consider the 
spin transport equation (\ref{spintrans}), which in the present case reads
\begin{equation}
\dot{d}(\vecp,\vecx,t) + \ui\,C(\Phi^t(\vecp,\vecx))\,\si_j\, 
d(\vecp,\vecx,t) = 0\ , \qquad d(\vecp,\vecx,0) = \eins_2 \ ,
\end{equation}
and is solved by the expression
\begin{equation}
\label{dsol}
d(\vecp,\vecx,t) = \cos(\al(\vecp,\vecx,t))\,\eins_2 - \ui\,
\sin(\al(\vecp,\vecx,t))\,\si_j\ ,
\end{equation}
where
\begin{equation}
\al(\vecp,\vecx,t) = \int_0^t C(\Phi^s(\vecp,\vecx))\ \ud s \ .
\end{equation}
The explicit solution (\ref{dsol}) of the spin transport equation demonstrates
that, even if the flow $\Phi^t$ on the base manifold $\Om_E$ is ergodic,
its $\SU(2)$-extension $Y^t$ cannot be ergodic on the product phase space
$\cM$. This is due to the fact that with (\ref{dsol}) one only explores a
one dimensional submanifold of the three dimensional group manifold of
$\SU(2)$. We therefore conclude that in the case of Pauli Hamiltonians 
ergodicity of $\Phi^t$ alone is not a sufficient criterion for quantum 
ergodicity.

\vspace*{0.5cm}
\subsection*{Acknowledgment}
We would like to thank Stefan Keppeler and Roman Schubert for useful 
discussions.

\providecommand{\bysame}{\leavevmode\hbox to3em{\hrulefill}\thinspace}

\end{document}